\documentclass[superscriptaddress,showpacs,twocolumn,aps]{revtex4}
\usepackage{graphicx}

\begin{document}
\input epsf

\title{Aluminium Nanowires: Influence of Work Hardening on
Conductance Histograms}
\author{I.K. Yanson}
\affiliation{B.I. Verkin Institute for Low Temperature Physics and
Engineering, National Academy of Sciences of Ukraine, 47, Lenin
Ave., 61103, Kharkov, Ukraine.} \affiliation{Kamerlingh Onnes
Laboratorium, Universiteit Leiden, Postbus 9504, 2300 RA Leiden,
The Netherlands.}
\author{O.I. Shklyarevskii}
\affiliation{B.I. Verkin Institute for Low Temperature Physics and
Engineering, National Academy of Sciences of Ukraine, 47, Lenin
Ave., 61103, Kharkov, Ukraine.} \affiliation{Institute for
Molecules and Materials, University of Nijmegen, Toernooiveld 1,
6525 ED Nijmegen, The Netherlands}
\author{J. M. van Ruitenbeek}
\affiliation{Kamerlingh Onnes Laboratorium, Universiteit Leiden,
Postbus 9504, 2300 RA Leiden, The Netherlands.}
\author{S. Speller}
\affiliation{Institute for Molecules and Materials, University of
Nijmegen, Toernooiveld 1, 6525 ED Nijmegen, The Netherlands}

\begin{abstract}
Conductance histograms of work-hardened Al show a series up to 11
equidistant peaks with a period of 1.15$\,\pm\,$0.02 of the
quantum conductance unit $G_0 = 2e^2/h$. Assuming the peaks
originate from atomic discreteness, this agrees with the value of
1.16 $G_0$ per atom obtained in numerical calculations by Hasmy
{\it et al.}.
\end{abstract}

\maketitle

Stability and self-organization phenomena in metallic nanowires
(NW) are controlled by an intimate combination of the quantum
nature of the conduction electrons and the atomic-scale surface
energy (for a review see Ref.~\onlinecite{Rev}). At the scale of
single atoms  Au, Pt, and Ir spontaneously form into
chains.\cite{Ohn,AYan1,Smi} For larger diameters more complex
morphologies including spiral and helical multi-shell structures
(often referred to as `weird wires') were predicted
\cite{Gul,Tosatti} and eventually observed in high-resolution
transmission electron microscopy (TEM).\cite{Kon,Osh,Osh1}

Electronic shell effects can stabilize and therefore favor certain
geometries. Independent series of stable nanowire diameters with
periodicities proportional to the square root of the conductance,
$\sqrt{G}$,  were observed for the free-electron-like alkali
metals \cite{AYan2,AYan3} and the noble metals.\cite{Med,Mar,Mar1}
Although these NWs were not imaged as by TEM, their stability was
deduced from the statistical analysis of frequently occurring
stable conductance values during breaking of the contacts.

With increasing NW diameter the contribution of the surface energy
becomes more important and stable configurations are governed by
the atomic packing.  This gives rise to an atomic shell filling
series and the crossover between electronic and atomic shell
structure has been observed in the conductance of alkali and noble
metals nanowires \cite{Mar,Mar1,AYan4} as  a change in the regular
period on the scale $\sqrt{G}$. Atomic shell filling has been
observed in TEM as exceptionally long and stable
wires,\cite{Kon1,Kiz,Rod} predominantly along $\left< 110\right>$.

During last two years the effects of pseudoelastic deformation and
shape memory in nanowires of the simple fcc metals (Ag, Cu) has
drawn considerable attention.\cite{park,liang} These effects occur
through the formation of defect-free twins during the
$\left<100\right>$ to $\left<110\right>/\{111\}$ reorientation
process. One can expect the emerging of stable configurations
during a cycle of contraction and elongation of the NW. However,
such configurations will be nanowire-specific and depend on NW
diameter, crystallographic orientation, existence of defects etc.
An example of striking repetitiveness of the conductance during
the cyclical deformation (without breaking) of a long Au neck at
4.2 K was presented already in Ref.~\onlinecite{unt}.

Recently, we have reported the observation of a new series of
structures periodic in $G$, rather than $\sqrt{G} $, in
conductance histograms of work hardened gold.\cite{Yan} We
suggested that heavy work hardening of the starting wire results
in a high density of inter-crystalline boundaries and leads to
texture of the most dense crystal planes [111], [100] and [110]
being perpendicular to the nanowire axis. The distances $\Delta g$
between the maxima in conductance histograms correspond to the
changes in the smallest cross section of densest planes by an
integer number of atoms. The increment in conductance is then
expected to scale as $\Delta g_{111}$/$\Delta g_{100}$/$\Delta
g_{110}$ $\cong$ 0.87/1/1.41 based on the size of the atomic unit
cells. The Fourier transform of the data revealed three
`frequencies' at a close ratio: $\sim0.8$, 1.1, and
$1.4\,G^{-1}_0$. The mechanical properties of nanocrystalline
materials (including NWs \cite{bietsch}) differ drastically from
those of single crystals. In case of work hardened materials the
mobility of atoms at grain boundaries may be high enough to adjust
the local orientation of the grains on each side of the contact
along one of the principal crystallographic axis.

Numerical simulations of conductance histograms for gold using a
parameterized tight-binding approach were performed by Dreher {\it
et al}.\cite{Dre} They demonstrated for the main crystallographic
orientations a series of peaks in histograms for the minimal cross
sections that is related to the atomic discreteness. However they
found this periodic structure in the minimal cross section was not
reflected in conductance histograms in their calculations.

Here we report the observation of atomic-size oscillations for
aluminium nanocontacts. The extension of this effect, previously
observed for noble metals, to trivalent Al is of principal
significance. One may anticipate that changing the nanowire cross
section by a single atom for noble metals results in a change in
conductance by approximately one quantum.\cite{Sch} In Al three
valence electrons give rise to three quantum channels: a highly
transmissive channel with $sp_z$ character and two low
transmissive channels with a $p_x, p_y$
character.\cite{Cuevas,Agr,Sch1} The most unexpected result of our
experiments is that the incremental conductance between adjacent
maxima in conductance histograms of work-hardened Al is again
close to one quantum conductance for up to about 11 oscillations
(see below). In case of Al we can no longer apply quasi-classical
considerations as we did for Au, and have to use a quantum
mechanical approach. Such approach was elaborated by Hasmy {\it et
al.},\cite{Has1,Has2} who performed an embedded atom molecular
dynamics method coupled with full quantum calculations of electron
transport using a procedure based on the {\it ab initio} Gaussian
embedded-cluster method.\cite{Pal} The results reveal a
statistically linear relationship between the conductance and the
number of aluminium atoms in the contact cross-section with the
slope equal to 1.16 $G_0$.

\begin{figure}[!b]
\includegraphics[width=7.9cm,angle=0]{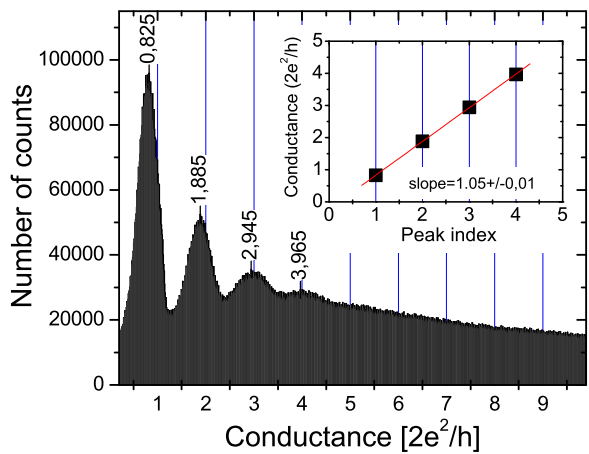}
\caption {Conductance histogram for annealed Al. Bias voltage 80
mV, number of traces 11900, $T=4.2$K. Inset:  The position of the
maxima in the conductance histogram versus peak index, giving a
slope of 1.05$\pm$0.01 $G_0$.} \label{fig1}
\end{figure}

Conductance histograms are constructed  from a large number of
contact-breaking traces and show the probability for observing a
given conductance value $G$. While for chemically inert gold
conductance histograms of reasonable quality can be measured in
air using very simple table-top devices\cite{Hansen} this is much
less obvious for chemically active aluminium, although some
results have recently been reported.\cite{ancuta} The main problem
is that the hardness of aluminium oxide exceeds by far that of Al
(9 versus 2.75 at Mohs scale of minerals hardness or 400 compared
to 8  at absolute hardness scale ). The direct electrical contact
between the electrodes can be established only by a `hard
indenting' procedure \cite{Sakai} causing substantial damage to
the lattice. An obvious solution to this problem is employing the
mechanically controllable break junction (MCBJ) technique,
described in details in Ref.~\onlinecite{Rev}. In our experiments
the notched aluminium wire is broken under cryogenic vacuum
conditions at 4.2 K exposing two electrodes with atomically clean
surfaces. Accurately controlled indentation-retraction cycles can
be obtained through the action of a piezoelectric element. On rare
occasions ($\leq 5\%$) featureless or ill-reproducible histograms
were observed with a small number of maxima positioned at
significantly lower $G$ as compare to the rest of the data.
Possibly this is caused by the break of the work hardened wire
occurring along an oxidized microcrack or grain boundary.

For well-annealed\cite{anneal} Al one typically observes several
peaks in the histogram near $G=nG_0$, with $n=1, 2, 3, 4$ (Fig.
1). One more peak is sometimes found near 5$\,G_0$, but in most
cases only peaks up to 3$\,G_0$ are visible. Although a
contribution of conductance quantization effects cannot be
excluded completely, the main reason of the appearance of the
peaks is the discreteness of the atomic structure of the
contacts.\cite{Rev,AYan5} The first peak results from single-atom
contacts and higher peaks are due to stable atomic arrangements
involving more atoms in the contact cross section (see Fig.~3 in
Ref.~\onlinecite{Has2}).

\begin{figure}[!t]
\includegraphics[width=7.9cm,angle=0]{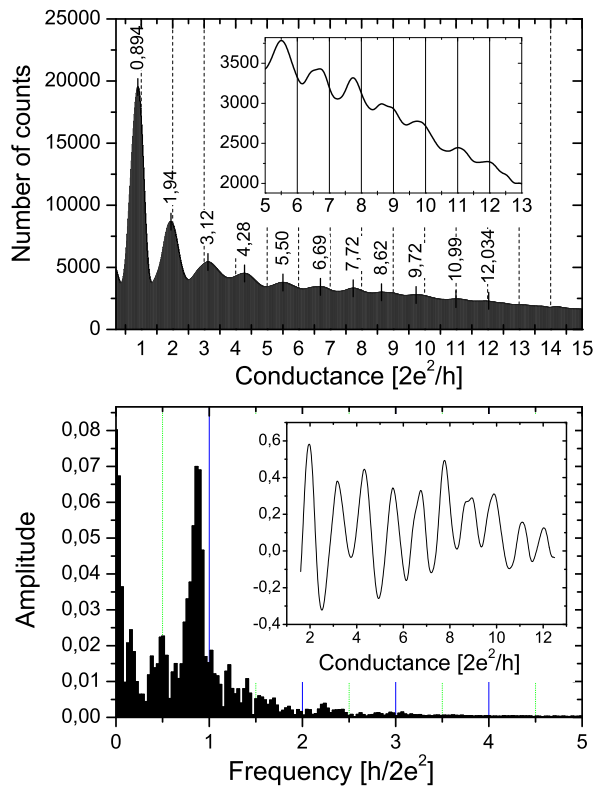}
\caption {Upper panel: Histogram exhibiting one of the highest
numbers of peaks for work hardened Al. Bias voltage 80 mV, number
of traces 21030, $T=4.2$K. Inset: Magnified portion of the
histogram between $G_0=5$--13. Lower panel: Fourier spectrum of
the conductance histogram presented above, after re-scaling as
shown in the inset (see text).  } \label{fig2}
\end{figure}

In our experiments we used two types of Al wires. Part of the
measurements was done with commercial `temper hard'
samples.\cite{Adv} Practically the same results were obtained by
work hardening of annealed wires. To this end we pulled wires
through a series of sapphire dies, reducing the diameter from 250
to 100$\mu$m. In both cases it was possible to observe a structure
periodic in $G$ having up to 11 oscillations in the conductance
range from 0 to 13 $G_0$ (Fig.~2, upper panel). The position of
maxima in the conductance histograms and their relative intensity
were accurately  reproduced for different samples.

The oscillating part of the histogram is presented in the inset to
the lower panel of Fig.~2. It was obtained by subtracting the
nearly linear background and by normalizing through division by
the enveloping curve. The Fourier spectrum of this re-scaled
histogram shows a peak at the `frequency' 0.87$\pm 0.01\,G_0^{-1}$
(lower panel in Fig.~2). Assuming the oscillations result from
atomic discreteness in analogy to those observed\cite{Yan} for Au,
this number is very close to the slope 0.86~$G_0^{-1}$ obtained in
numerical simulations for nanowires with a $\left<111\right>$
oriented axis.\cite{Has2}

The histogram presented above was obtained for conductance traces
recorded while breaking the contact. One can also take histograms
for traces of contact-closing, which we refer to as return
histograms. We measured direct and return histograms for
work-hardened Al by applying an isosceles triangle ramp voltage to
the piezodriver and separating data for the pulling and pushing
part of the cycle. Surprisingly these histograms are practically
equivalent (Fig.~3). This is quite different from our observations
for work-hardened gold, where the number of the atomic-scale
oscillations was greatly reduced in the return histograms and the
electronic shell structure was observed in 40\% of the
latter.\cite{Yan}

\begin{figure}[!t]
\includegraphics[width=7.9cm,angle=0]{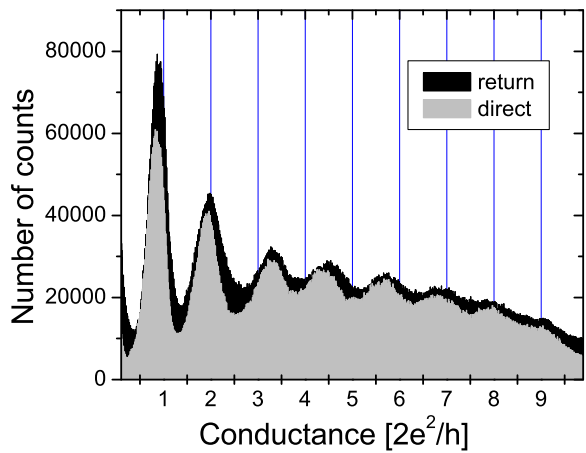}
\caption {Direct conductance histogram (obtained for breaking
contacts, gray) versus return histogram (closing contacts, black)
for the same cycle of measurement.} \label{fig3}
\end{figure}

Conductance histograms for work-hardened Au invariably show a
clear `beating' pattern in the peak intensities related to the
existence of several closely separated periodicities. In contrast
for Al histograms with the largest number of peaks (10--11) we
observed only a smooth decrease of the enveloping curve and a
single frequency in the Fourier spectra around 0.87$\,G_0$. In
search for different periods we paid attention to histograms with
lower numbers of peaks (Fig.~4). One notices kinks in the envelope
curve for the histogram maxima (inset in Fig.~4). Although these
irregularities may indicate contributions from the various axis
orientations, the magnitude of the effect is rather small and the
number of peaks in the histogram is not sufficient for more
detailed analysis.

\begin{figure}[!b]
\includegraphics[width=7.9cm,angle=0]{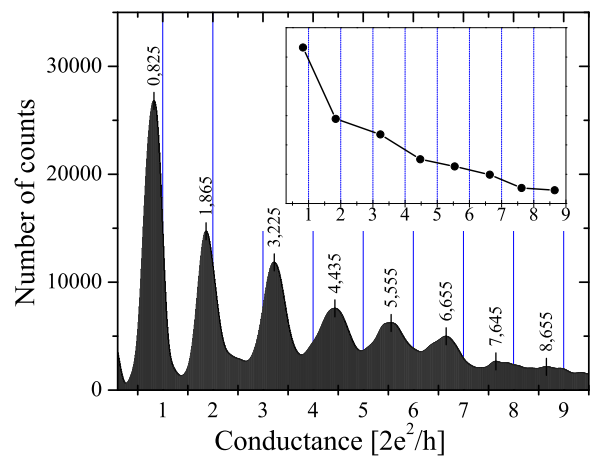}
\caption {Upper panel. Typical conductance histogram  for work
hardened Al with 8 maxima. Inset: enveloping curve for histogram.
Bias voltage 99 mV, number of curves 11196, T=4.2 K.} \label{fig4}
\end{figure}

Assuming the $\left<111\right>$ periodicity being due to atom-size
oscillation effects similar to our previous results on Au
wires\cite{Yan} we can estimate a Fermi wave vector $k_{\rm F}$ in
Al for effective free electrons propagating along (111) plane per
one atom change of (111) cross section. Using the Sharvin formula,
we obtain a value of $k_{\rm F}=1.55\cdot 10^{10}$øm$^{-1}$ for
the effective momentum in Al.

The picture emerging from our experiments on Au \cite{Yan} and
more specifically for Al is as follows. For single crystalline
necks with the highest atomic packing in the transverse crystal
plane (like the (111) plane) not only are the Al atoms stacked at
definite positions in the nodes of the lattice but their orbitals
are also oriented in the same way.  Hence three valence electrons
contribute to the conductance with a contribution that scales with
the number of atoms. This property is at the heart of the observed
linear dependencies of contact conductance on the numbers of atoms
in the smallest cross section of the nanowire. These atomic
arrangements are associated with transient minima of the lattice
free energy leading to maxima on conductance histograms.

Summarizing, we observed a new periodic structure in the
conductance histograms of work hardened Al wires, which agrees
with a series of stable wire cross sections given by integer
numbers of atoms, combined with a conductance per atom as
calculated by Hasmy {\it et al.}\cite{Has2} for the (111)
orientation of nanowire axis. The experimentally observed peak in
the Fourier transform at 0.87$G_0^{-1}$ is very close to the
calculated value 0.86$G_0^{-1}$. We believe that this periodic
structure is due to the same mechanism suggested for Au
atomic-scale oscillations observed earlier.\cite{Yan}

Part of this work was supported by Nanotechnology network in the
Netherlands NanoNed, the Dutch nanotechnology programme of the
Ministry of Economic Affairs, the Stichting voor Fundamenteel
Onderzoek der Materie (FOM) which is financially supported by the
Netherlands Organization for Scientific research (NWO), and the
NANO-programm of Ukraine. O.I.S. wishes to acknowledge the FOM for
a visitor's grant.


\begin{thebibliography}{0}
\expandafter\ifx\csname natexlab\endcsname\relax\def\natexlab#1{#1}\fi
\expandafter\ifx\csname bibnamefont\endcsname\relax
  \def\bibnamefont#1{#1}\fi
\expandafter\ifx\csname bibfnamefont\endcsname\relax
  \def\bibfnamefont#1{#1}\fi
\expandafter\ifx\csname citenamefont\endcsname\relax
  \def\citenamefont#1{#1}\fi
\expandafter\ifx\csname url\endcsname\relax
  \def\url#1{\texttt{#1}}\fi
\expandafter\ifx\csname urlprefix\endcsname\relax\def\urlprefix{URL }\fi
\providecommand{\bibinfo}[2]{#2}
\providecommand{\eprint}[2][]{\url{#2}}

\end{thebibliography}


\begin{thebibliography}{99}

\bibitem{Rev} N. Agra\"{\i}t, A. Levy Yeyati, and J. M. van Ruitenbeek,
Phys. Rep. {\bf 377}, 81 (2003).
\bibitem{Ohn} H. Ohnishi, Y. Kondo, and K. Takayanagi, Nature (London) {\bf 395}, 780 (1998).
\bibitem{AYan1} A. I. Yanson, G. Rubio Bollinger, H. E. van den Brom,
N. Agra\"{\i}t, and J. M. van Ruitenbeek, Nature (London) {\bf
395}, 783 (1998).
\bibitem{Smi} R. H. M. Smit, C. Untiedt, A. I. Yanson, and J. M.
van Ruitenbeek, Phys. Rev. Lett. {\bf 87}, 266102 (2001).
\bibitem{Gul}  O. Gulseren, F. Ercolessi, and E. Tosatti, Phys. Rev.
Lett. {\bf 80}, 3775 (1998).
\bibitem{Tosatti} E. Tosatti and S. Prestipino, Science, {\bf 289}, 561
(2000).
\bibitem{Kon}  Y. Kondo and K. Takayanagi, Science {\bf 289}, 606 (2000).
\bibitem{Osh} Y. Oshima, H. Koizumi, K. Mouri, H. Hirayama, K.
Takayanagi, and Y. Kondo, Phys. Rev. B {\bf 65}, 121401 (2002).
\bibitem{Osh1} Y. Oshima, A. Onga, and K. Takayanagi, Phys. Rev.
Lett. {\bf 91}, 205503 (2003).
\bibitem{AYan2}  A. I. Yanson, I. K. Yanson, and J. M. van Ruitenbeek,
Nature (London) {\bf 400}, 144 (1999).
\bibitem{AYan3}  A. I. Yanson, I. K. Yanson, and J. M. van Ruitenbeek,
Phys. Rev. Lett. {\bf 84}, 5832 (2000).
\bibitem{Med}  E. Medina, M. Diaz, N. Leon, C. Guerrero, A. Hasmy,
P. A. Serena, and J. L. Costa-Kramer, Phys. Rev. Lett. {\bf 91},
026802 (2003).
\bibitem{Mar} A. I. Mares, A. F. Otte, L. G. Soukiassian, R. H. M. Smit,
and J. M. van Ruitenbeek, Phys. Rev. B {\bf 70}, 073401 (2004).
\bibitem{Mar1} A. I. Mares and J. M. van Ruitenbeek, Phys. Rev. B {\bf 72},
205402 (2005).
\bibitem{Kon1}  Y. Kondo and K. Takayanagi, Phys. Rev. Lett. {\bf 79}, 3455
(1997).
\bibitem{Kiz} T. Kizuka, Phys. Rev. Lett. {\bf 81}, 4448 (1998).
\bibitem{Rod} V. Rodrigues, T. Fuhrer, and D. Ugarte, Phys. Rev.
Lett. {\bf 85}, 4124 (2000).
\bibitem{park} H.S. Park, K. Gall, and J.A. Zimmerman, Phys.Rev.Lett.
{\bf 95}, 255504 (2005).
\bibitem{liang} W. Liang and M. Zhou, Phys.Rev.B {\bf 73}, 115409 (2006).
\bibitem{unt} C. Untiedt, G. Rubio, S. Vieira, and N. Agra\"{\i}t,
Phys. Rev. B {\bf 56}, 2154 (1997).
\bibitem{AYan4}  A. I. Yanson, I. K. Yanson, and J. M. van Ruitenbeek,
Phys. Rev. Lett. {\bf 87}, 216805 (2001).
\bibitem{Yan} I. K. Yanson, O. I. Shklyarevskii, Sz. Csonka, H. van Kempen, S. Speller,
A. I. Yanson, and J. M. Van Ruitenbeek, Phys. Rev. Lett. {\bf 95},
256806 (2005).
\bibitem{bietsch} A. Bietsch and B. Michel, \apl {\bf 80} 3346 (2002).
\bibitem{Dre} M. Dreher, F. Pauly, J. Heurich, J. C. Cuevas, E. Scheer,
and P. Nielaba, Phys. Rev. B {\bf 72}, 075435 (2005).
\bibitem{Sch}  E. Scheer, W. Belzig, Y. Naveh, M.H. Devoret, D. Esteve,
and C. Urbina, \prl {\bf 86}, 284 (2001).
\bibitem{Cuevas} J. C. Cuevas, A. Levy Yeyati, and A. Mart\'{\i}n-Rodero,
\prl {\bf 80}, 1066 (1998).
\bibitem{Agr} J. C. Cuevas, A. Levy Yeyati, A. Mart\'{\i}n-Rodero, G. Rubio Bollinger,
C. Untiedt, and N. Agra\"{\i}t, \prl {\bf 81}, 2990 (1998).
\bibitem{Sch1} E. Scheer, P. Joyez, D. Esteve, C. Urbina, and M.H.
Devoret, Phys. Rev. Lett., {\bf 78}, 3535 (1997).
\bibitem{Has1}  A. Hasmy, E. Medina, and P. A. Serena, Phys. Rev. Lett. {\bf 86}, 5574 (2001).
\bibitem{Has2} A. Hasmy, A. J. Perez-Jimenez, J. J. Palacios, P. Garcia-Mochales,
J. L. Costa-Kramer, M. Diaz, E. Medina, and P. A. Serena, Phys.
Rev. B {\bf 72}, 245405 (2005).
\bibitem{Pal}  J.J. Palacios, A.J. Perez-Jimenez, E. Louis, E. SanFabian,
and J.A. Verges, \prb {\bf 66}, 035322 (2002).
\bibitem{Hansen} K. Hansen, E. L{\ae}gsgaard, I. Stensgaard, and F. Besenbacher,
\prb {\bf 56}, 2208 (1997).
\bibitem{ancuta} A.I. Mares, D.F. Urban, J. B{\" u}rki, H.
Grabert, C.A. Stafford, and J.M. van Ruitenbeek, cond-mat/0703589,
Nanotechnology, in print.
\bibitem{Sakai} J. Mizobata, A. Fujii, S. Kurokawa, and A. Sakai,
\prb {\bf 68}, 155428 (2003)
\bibitem{AYan5}  A.I. Yanson and J.M. van Ruitenbeek, \prl {\bf 79}, 2157 (1997).
\bibitem{anneal} We used aluminium wire 125 $\mu$ diameter, ``temper as drawn'',
99.99\%+ , Goodfellow Corp. annealed in high vacuum at 350 C for 72
hours.
\bibitem{Adv} Aluminium wire, 125 $\mu$ diameter, ``temper hard'', 99.99\%, Advent
Research Materials Ltd., Catalogue No. AL500111

\end{thebibliography}
\end{document}